\documentclass[prl,showpacs,twocolumn]{revtex4}

\usepackage{graphicx}

\begin{document}

\title{Spin polarized electron transport near the Si/SiO$_2$ interface}

\author{Hyuk-Jae Jang}
\author{Ian Appelbaum}
\altaffiliation{appelbaum@physics.umd.edu}
\affiliation{Center for Nanophysics and Advanced Materials and Department of Physics, University of Maryland, College Park MD 20742 USA}

\begin{abstract}
Using long-distance lateral devices, spin transport near the interface of Si and its native oxide (SiO$_2$) is studied by spin-valve measurements in an in-plane magnetic field and spin precession measurements in a perpendicular magnetic field at 60K. As electrons are attracted to the interface by an electrostatic gate, we observe shorter average spin transit times and an increase in spin coherence, despite a reduction in total spin polarization. This behavior, which is in contrast to the expected exponential depolarization seen in bulk transport devices, is explained using a transform method to recover the empirical spin current transit-time distribution and a simple two-stage drift-diffusion model. We identify strong interface-induced spin depolarization (reducing the spin lifetime by over two orders of magnitude from its bulk transport value) as the consistent cause of these phenomena.  
\end{abstract}


\maketitle

Recent advances in the development of techniques for electrical injection and detection of spin-polarized electrons have revolutionized semiconductor spintronics.\cite{APS} Among the materials studied so far, silicon (Si) stands out due to its weak spin orbit and hyperfine effects,\cite{ZUTICNATURE, IGORSIPRL} effectively insulating electron spin from relaxation mechanisms and leading to long spin lifetimes (e.g. $>500$ns at 60K) in bulk.\cite{BIQINPRL} Although this long lifetime was subsequently exploited to demonstrate long distance spin transport of over 2mm in quasi-lateral devices,\cite{2MM} the transport mode was still largely bulk in character, where the inversion symmetry of the diamond lattice potential protects spin degeneracy of the conduction band and suppresses spin-orbit effects. 

At interfaces where bulk transitions to 2-dimensional transport, however, the lattice inversion symmetry is broken. This opens the possibility of the emergence of (weak) spin orbit effects including Bychkov-Rashba spin manipulation.\cite{NITTA} Besides the bare surface of Si, which is prone to contamination and is difficult to electrostatically gate,\cite{KANE} the buried Si/SiO$_2$ interface is of interest in exploring this physics experimentally because of its high conduction band offset with the insulating oxide and its technological importance in charge-based Si field-effect devices.

Although it is widely expected that the approach to two-dimensional transport tends to preserve spin polarization as a result of limiting orbital degrees of freedom,\cite{KUNIHASHI} here we show (using gate-tunable lateral Si spin transport devices and a corresponding transport theory) that attracting bulk conduction electrons to the thermally-grown Si/SiO$_2$ interface causes a massive spin lifetime reduction. These results are important for the development of gated semiconductor spin transport devices, such as the canonical Datta-Das transistor,\cite{DATTADAS} and may have a similar physical origin to effects seen in spin transport in graphene/SiO$_2$ devices, where the observed spin lifetime is many orders of magnitude lower than expected.\cite{VANWEES}

\begin{figure}
\includegraphics[scale=0.425]{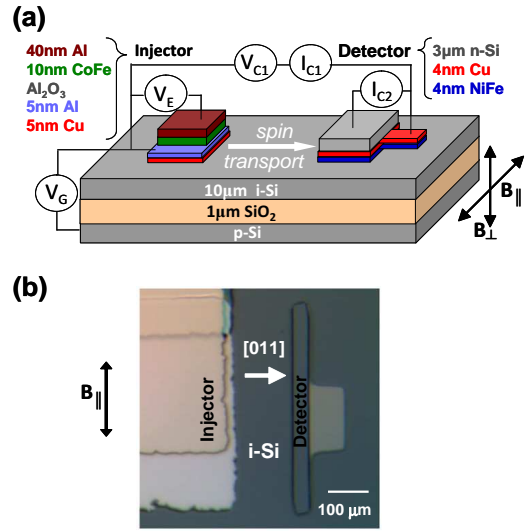}
\caption{(a) Schematic view of the electrostatically-gated lateral Si spin transport devices, showing electrical connection configuration. (b) Plan-view micrograph of a typical device.\label{FIG1}}
\end{figure}

Our lateral spin transport devices utilize ballistic hot-electron spin injection and detection techniques, similar to previously-reported vertical-transport devices.\cite{APPELBAUMNATURE, BIQINPRL, DEPHASING, 2MM, SPINFETEXPT, BIQINJAP, 35PERCENT, DOPED} However, the lateral geometry used here required modified fabrication methods. First, to make a semiconductor-metal-semiconductor spin detector, ultra-high vacuum metal film wafer bonding was performed with two different silicon-on insulator (SOI) wafers using a thermally evaporated Ni$_{80}$Fe$_{20}$ (4nm)/Cu (4nm) bilayer; one SOI wafer consists of 3$\mu$m single-crystal (100) n-type silicon for the detector hot electron collector and the other of 10$\mu$m single-crystal (100) undoped silicon for the transport channel. The rather large thickness of 10$\mu$m was chosen to achieve wide proximity control of spin-polarized conduction electrons to the Si/SiO$_2$ interface below it using the p-type Si handle substrate wafer as an electrostatic gate. Conventional wet-etching methods, along with photolithography, were used to pattern the spin detector and the subsequently exposed undoped silicon layer was Ar$^+$ ion-milled to eliminate a conductive silicide on the surface, which could be detrimental to spin transport.\cite{35PERCENT} On the exposed undoped silicon layer, a hot-electron spin injector was assembled with a ferromagnetic cathode tunnel junction having a structure of Al (40nm)/Co$_{84}$Fe$_{16}$ (10nm)/Al$_{2}$O$_{3}$/Al (5nm)/Cu (5nm)/undoped-Si (10 $\mu$m). A schematic diagram of the completed lateral device structure and a plan-view micrograph are shown in Figs. \ref{FIG1}(a) and (b), respectively. The distance from the edge of the base (the Al/Cu metal layer under Al$_{2}$O$_{3}$) to the edge of the detector is approximately 150 $\mu$m, but the edge of the emitter tunnel junction is receded from the base edge by approximately 20$\mu$m. XRD measurements confirmed that electron spin transport from the injector to detector is along the [011] crystalline axis.

To perform electron spin transport measurements, a tunnel junction emitter voltage V$_{E}$ larger than the Cu/i-Si injector Schottky barrier height is applied to inject spin-polarized hot electrons ballistically into the 10 $\mu$m-thick undoped silicon channel. Because the lateral electric field caused by the accelerating voltage drop V$_{C1}$ is screened by the equipotential injector base layer, these electrons first must diffuse to the channel region and are then carried primarily by drift toward the detector. After coupling to hot electron states in the Ni$_{80}$Fe$_{20}$, the ballistic component of the charge current is filtered by the asymmetry between spin-up and spin-down mean-free-paths in the ferromagnet, so that the number of electrons coupling to conduction band states in the n-type Si detection collector (comprising the signal current $I_{C2}$) will indicate the final spin polarization of the transported electrons. Low operating temperatures are required to freeze out unpolarized thermionic currents over the Schottky barriers, which would otherwise dilute the small ballistically-injected spin polarized current.

\begin{figure}
\includegraphics[width=8cm, height=7.25cm]{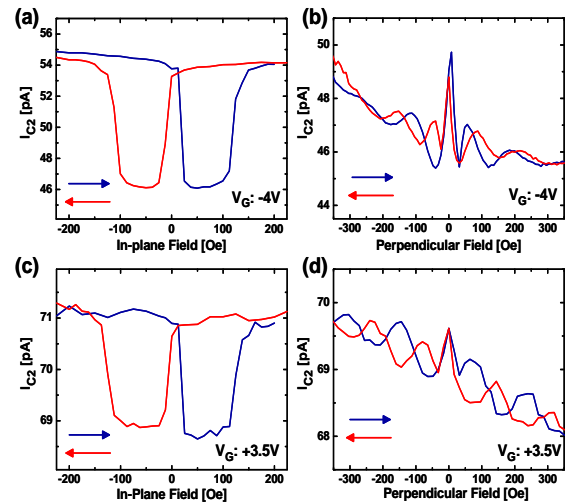}
\caption{(a) Spin-valve measurement in an in-plane magnetic field and (b) spin-precession (Hanle) measurement in a perpendicular magnetic field under negative gate biasing conditions which push transported electrons away from the Si/SiO$_2$ interface. (c) Spin valve measurement and (d) spin precession measurement under positive gate biasing conditions which pull transported electrons toward the Si/SiO$_2$ interface. All measurements use emitter voltage $V_E=$-1.3V and accelerating voltage $V_{C1}=$20V at 60K. Arrows indicate direction of magnetic field sweep for the appropriately colored data. \label{FIG2}}
\end{figure}

Two types of measurements were performed using external magnetic field sweeps, first in an in-plane geometry and second in a perpendicular field geometry at 60K with constant $V_E=$-1.3V. The resulting spin-valve and spin-precession data at constant $V_{C1}$=20V are shown in Fig. \ref{FIG2} for gate voltage bias $V_G$=-4V, where injected electrons are pushed away from the Si/SiO$_2$ interface, ((a) and (b), respectively), and $V_G$=+3.5V, where they are attracted to the interface, ((c) and (d), respectively). 

A comparison of the spin-valve data in Fig. \ref{FIG2}(a) and (c) reveals similar device response: As the external magnetic field sweeps through the in-plane coercive field of the magnetically softer Ni$_{80}$Fe$_{20}$, an antiparallel injector-detector magnetization configuration is obtained, and the signal current determined by projection of spin orientation on detector magnetization $I_{C2}$ is reduced. This is due to a decrease in ballistic mean-free-path in the detector ferromagnet, and is identical to the behavior of vertical spin transport devices utilizing the same detection technique.\cite{APPELBAUMNATURE, BIQINPRL, DEPHASING, 2MM, SPINFETEXPT, BIQINJAP, 35PERCENT, DOPED} However, marked qualitative differences are clear from a comparison of the spin precession data in a perpendicular magnetic field, shown in Figs. \ref{FIG2}(b) and (d): although the magnitude of spin precession oscillation is reduced as the gate voltage is varied from negative in Fig. \ref{FIG2}(b) to positive in Fig. \ref{FIG2}(d), the coherence (e.g. the number of clear oscillations) is dramatically increased. Furthermore, the magnetic field period of these coherent oscillations is larger in Fig. \ref{FIG2}(d). 

Since these oscillations are due to spin precession at angular frequency $\omega=g \mu_B B/ \hbar$ (where $g$ is the electron spin g-factor, $\mu_B$ is the Bohr magneton, $B$ is the magnetic field, and $\hbar$ is the reduced Planck constant) during transport time $t$, the period of the oscillations $B_{2\pi}$, caused by a change in the average spin orientation, can be used to recover $t=2 \pi \hbar /g \mu_B B_{2\pi}$.\cite{BIQINJAP} Therefore, the larger oscillation period seen in Fig. \ref{FIG2}(d) indicates a shorter transport time, despite the identical voltage drop and drift field between injector and detector.

This behavior is in contrast to vertical Si spin transport devices, where shorter spin transport time leads to an {\it {increase}} in observed spin polarization due to a simple exponential spin decay mechanism.\cite{BIQINPRL} However, the geometry here is very different; for example, the lateral extent of the injector ($\approx 400 \mu m$) and detector ($\approx 50 \mu m$) is expected to cause an unavoidable systematic source of decoherence (or spin ``dephasing'') because of a transit-length uncertainty leading to transit time uncertainty $\Delta t$ and suppression of oscillations due to spin precession angle uncertainty $\Delta \theta=\omega \Delta t$.\cite{LOU} In light of this expected geometrical dephasing, the many coherent oscillations seen in Fig. \ref{FIG2}(d) is surprising.

One simple explanation for this observed behavior could be a spin lifetime on the order of the transit time $t$. This would suppress contributions to the magnetic field-dependent spin signal by electrons which arrive at the detector after long transit times, shortening the observed average transit time and the overall polarization, and decrease the width of the distribution and therefore the dephasing. However, previous studies\cite{BIQINPRL} have established spin lifetimes in bulk undoped Si using vertical transport devices at this temperature of at least 520 ns, and the apparent transit time here is $\approx$ 2ns, so a drastic reduction in spin lifetime induced by the presence of the interface would be necessary to support this explanation. 

Fortunately, we have empirical access to the spin current transit time distribution through the spin precession data. This signal is a weighted sum of $\cos{\omega t}$ contributions (due to precession at angular frequency $\omega$) from the projection of spin orientation onto the magnetization axis of the detector thin film.\cite{LOU, DEPHASING, BIQINPRL, SPINFETTHRY} The weighting function $P(t)$ is just the spin current transit time distribution at the detector so that our expected signal is given by:

\begin{equation}
I_{C2}\propto \int_0^\infty P(t)\cos{\omega t}dt.
\label{INTEGRAL1} 
\end{equation}

\noindent Because $P(t<0)=0$, we can extend the lower bound of integration to $-\infty$ in Eqn. \ref{INTEGRAL1} without consequence, and recover an expression equivalent to the real part of the Fourier transform of the spin current transit time distribution:

\begin{equation}
I_{C2}\propto Re\left[ \int_{-\infty}^\infty P(t)e^{i\omega t}dt \right].
\label{INTEGRAL2} 
\end{equation}

\begin{figure}
\includegraphics[width=9cm, height=5cm]{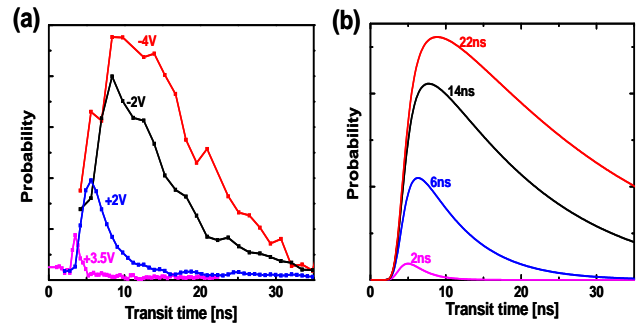}
\caption{(a) Spin current transit-time probability distributions for different gate voltages $V_G$ as labeled, derived from Fourier transforms of spin precession measurements using Eqn. \ref{INTEGRAL2} for $V_{C1}=$10V. Distributions are normalized by the total spin signal $I_{C2}$ in a magnetically parallel injector and detector configuration. (b) Spin transit-time distribution functions calculated using the two-regime drift-diffusion model described in the text for different spin lifetimes as labeled. \label{FIG3}}
\end{figure}

\noindent Since the spin precession signal $I_{C2}$ is real, the transform has even symmetry in $\omega$, so the real part of the Fourier transform of the spin precession data will then recover the spin current transit time distribution.  

When we perform this transformation, spin current transit time distributions such as those for $V_{C1}=$10V shown in Fig. \ref{FIG3}(a) are extracted. At a gate voltage bias of $V_G$=-4V, the distribution is wide and peaked near 10ns. However, as the gate voltage is made more positive and electrons are attracted toward the Si/SiO$_2$ interface, the resulting spin transit time distribution contracts to lower transit times but always remains within the wider distribution. This behavior clearly suggests a suppression of contributions from long transit-time electrons by fast depolarization at the Si/SiO$_2$ interface.  

To more quantitatively test this hypothesis, we simulate spin transit time distributions using a 1-dimensional model where injected spin-polarized electrons (screened from the lateral drift electric field by the injector base layer) diffuse at the Si/SiO$_2$ interface until they reach the edge of the base, where they then are carried by drift and diffusion to the detector. Their total transit-time distribution is then given by the convolution of the transit time distribution of the diffusion-dominated path under the injector base (where spins are injected a distance $w<x'<w+W$ from the edge of the emitter base, where $w$ is the exposed region of injector base under the emitter (here, 20 $\mu$m), and $W$ is the length of the base covered by the emitter (here, 400 $\mu$m)) with the drift-dominated transit time distribution for transport in the $L= 150 \mu$m-long channel between the injector and detector. The former (diffusion-only) distribution is given by a sum of all contributions (each the impulse response or ``Green's function'' of the spin diffusion equation\cite{YUFLATTE}) from along the wide injector emitter to the beginning of the drift-dominated region: 

\begin{equation}
p_1(t)=\int_{w}^{w+W}\frac{1}{2\sqrt{\pi Dt}}e^{-\frac{x'^2}{4Dt}}e^{-t/\tau}dx',
\end{equation}

\noindent where $D$ is the diffusion coefficient and $\tau$ is the spin lifetime. The latter (drift-diffusion) distribution is just the Green's function of the spin drift-diffusion equation with absorbing boundary conditions at the detector $x=L$:\cite{APPELBAUMREVIEW} 

\begin{equation}
p_2(t)=\frac{1}{2\sqrt{\pi Dt}}\left[e^{-\frac{(x-vt)^2}{4Dt}}-e^{\frac{Lv}{D}}e^{-\frac{(x-2L-vt)^2}{4Dt}}\right]e^{-t/\tau},
\label{DRIFTDIFF}
\end{equation}

\noindent where $v$ is the drift velocity. Neglecting any potential spin blockade effects\cite{PERSHIN}, the \emph{spin current} distribution is then proportional to the diffusion current at the absorbing detector boundary, $P(t)\propto -D\frac{d}{dx}(p_1\otimes p_2)|_{x=L}$. 
 
Simulated spin transit time distributions (for several spin lifetimes $\tau=$2ns, 6ns, 14ns, and 22ns) are shown in Fig. \ref{FIG3}(b) for comparison with the empirical distributions in Fig. \ref{FIG3}(a). The transport parameters (determined from fits of spin precession measurements from 350$\mu$m undoped-Si vertical spin-transport devices at 60K)\cite{DEPHASING} are mobility $\mu=$5833 cm$^2$/Vs and $D=$1500 cm$^2$/s. We do not intend an exact fit of the empirical distributions; rather, we show that this model captures the salient qualities such as the similar peak-probability arrival time shift, polarization reduction, and transit-time uncertainty ($\Delta t$) suppression with spin lifetime reduction. The details of the empirical distributions are likely difficult to fit with a 1-dimensional model like the one we have employed here, since the electrostatics in the Si transport channel will likely couple the otherwise independent voltage bias variables $V_{C1}$ and $V_G$.

\begin{figure}
\includegraphics[width=8cm, height=10cm]{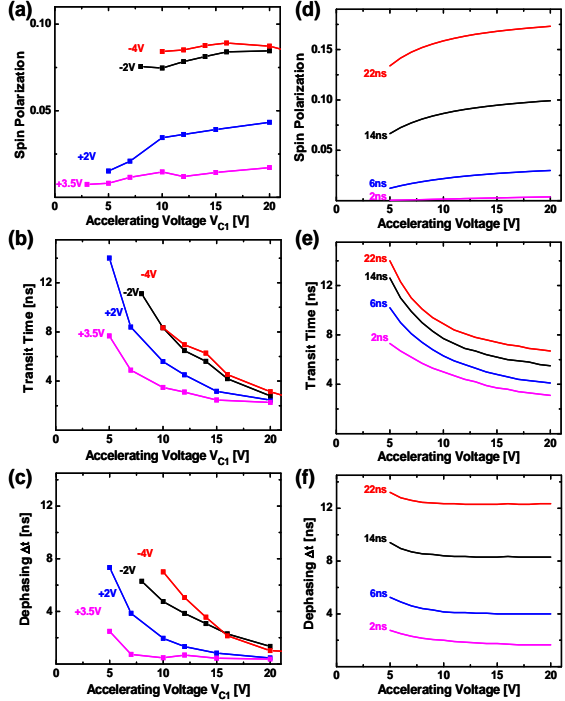}
\caption{Total spin polarization, most probable transit time, and dephasing (given by the half-width half-maximum), of the empirical spin transit time distributions ((a)-(c), respectively) and the simulated distributions ((d)-(f), respectively).  \label{FIG4}}
\end{figure}

Further support for the conclusion that the interface contributes to a reduction in spin lifetime is provided by a comparison of empirical and simulated spin transport distribution characteristics shown in Figs. \ref{FIG4}(a)-(c) and (d)-(f), respectively. As the drift electric field increases with increasing accelerating voltage bias $V_{C1}$, the empirical spin polarization saturates, transit time decreases, and spin dephasing (here, the distribution half-width at half maximum) decreases. At any constant $V_{C1}$, increasing $V_G$ tends to decrease all these parameters. Analysis of the simulated distributions shows a very similar behavior as shown in Figs. \ref{FIG4}(d)-(f). Remaining differences in the value of spin polarization between Figs. \ref{FIG4}(a) and (d) can be accounted for by a non-ideal spin polarization at the injection point, and the large-$V_{C1}$ separation between transit-time uncertainty (dephasing) seen in Fig. \ref{FIG4}(f) that is not seen in the experimental results in Fig. \ref{FIG4}(c) is likely due to the details of channel electrostatics as mentioned previously.

The microscopic cause of the observed severe spin lifetime suppression close to the interface is not clear. Interface roughness and presence of charged impurities are expected to enhance momentum scattering and increase spin depolarization proportionately via Elliott scattering,\cite{ELLIOTT} but the experimental spin transit distributions do not show a shift of the low-transit-time edge to longer times that would result from such a decrease in charge-transport mobility. However, the random locations of these charged impurities could cause a fluctuating Bychkov-Rashba field leading to an enhancement of spin relaxation.\cite{SHERMAN} The possibility also remains that the broken lattice inversion symmetry at the interface causes a D'yakonov-Perel'-type spin depolarization mechanism,\cite{DYAKONOV} and our future studies using these devices will explore whether this is the case, or if different effects such as localized spin exchange with paramagnetic defects in the oxide\cite{YABLONOVITCH} or dangling bonds at the interface\cite{MOODERA} are to blame. Regardless, this effect could be exploited for novel device operation requiring electronic control over spin lifetime,\cite{FLATTE} and could at least partially be to blame for an absence of evidence for spin injection in prior attempts at spin transport in Si.

We gratefully acknowledge B. Huang for assistance with fabrication and D. Kan for providing XRD measurements. This work was funded by DARPA/MTO, the Office of Naval Research, and the National Science Foundation.

\end{document}